\documentstyle[aps,prb]{revtex}
\draft
\begin{document}

\title{Kohn-Luttinger effect in nested Fermion liquids}
\author{Hyok-Jon Kwon\cite{Kwon}}
\address{Department of Physics, Brown University, Providence,
RI 02912-1843, USA}
\date{October 2, 1996}

\maketitle

\begin{abstract}
We study the Kohn-Luttinger effect in a two-dimensional (2D) nested Fermion
liquid with a repulsive interaction via the renormalization group 
method and identify the resulting order parameter symmetry.
Using the band structure of the 2D Hubbard model close to half-filling
as a prototype, we construct an effective low-energy theory.
We use multidimensional bosonization to incorporate the zero sound channel
and find marginal Fermi liquid behavior in the absence of any instability.
We show an analog of the Landau theorem in nested Fermion liquids,
which serves as the criterion of the BCS instability.
Including repulsive or antiferromagnetic exchange interactions in the
low-energy theory, we show that the $d_{x^2-y^2}$-wave BCS channel is
renormalized to be the most attractive. Below half-filling, when the
nesting is not perfect, there is competition between the spin-density-wave
(SDW) and the BCS channels; when the SDW coupling is small enough,
there occurs a $d_{x^2-y^2}$-wave superconducting instability at
sufficiently low temperatures.
\end{abstract}
{cond-mat/9610020}
\section{Introduction}
\label{sec:intro}

One question of recent interest concerns the possible
breakdown of the
Fermi liquid in the presence of
 singular or even regular interactions. One example is
the Luttinger liquid in one spatial dimension where the quasiparticle pole is
destroyed by a short-range interaction
although in this case the low-energy excitation remains gapless. The
existence of the Luttinger liquid in higher spatial dimensions has also been
proposed but is still controversial.\cite{Anderson}
 Another example is the Kohn-Luttinger
(KL) effect: A Fermi liquid state is unstable against one or more
BCS channels in the presence of a repulsive interaction.\cite{KL} It has been
shown that a short-range repulsive interaction can induce an attractive
BCS channel at large orbital angular momentum modes.
In three-dimensional Fermi gases,
the Kohn-Luttinger effect was found in the $p$-wave
BCS channel.\cite{Baranov}
This notion was extended to two-dimensional Fermi liquids
and a weaker KL effect was also found in the $p$-wave channel although
the effect is not visible in a second-order perturbative
expansion.\cite{Chubukov}

However, the KL effects in isotropic Fermi liquids are weak
because they are found in sub-leading corrections in
the renormalization group (RG) sense;
the resulting $T_c$ is expected to be low.
For example, in three dimensions, the KL correction to the $\beta $
function is found to
be $O(s^{-7/4})$ where $s$ is the dimensionless
length scale which is taken to be a
large number in the low-energy limit.\cite{Shankar}
But in the case of nested Fermion liquids,  the effect is more dramatic
since the nesting gives rise to a logarithmic flow in the repulsive
channel.
Baranov {\it et al.}\cite{Baranov} studied the KL effect in a
two-dimensional Hubbard model at low filling factors where
the Fermi surface is almost circular and found
an attractive channel with $d_{xy}$-wave symmetry, but it is also a
sub-leading effect which originates from the directional property
of the square lattice.
In this paper we  study the KL effect in two dimensions
in a nested square Fermi surface.

The subject of this paper may be of interest in the context of
the cuprate high-$T_c$ superconducting
materials. The KL effect implies that an initially
repulsive interaction between electrons can induce an effective
attractive BCS interaction among quasiparticles lying in the thin
momentum shell around the Fermi surface. Hence this is one possible
candidate for a non-phononic mechanism of cuprate
superconductivity.\cite{Ruvalds}
Indeed, we find that the repulsive interaction, which naturally contains
an antiferromagnetic (AF) exchange channel, favors a $d_{x^2-y^2}$-wave
BCS instability. This is in accordance with 
expectations.\cite{Monthoux,Ruvalds}

The low-energy theory of the two-dimensional Hubbard model
is complicated by several features of the anisotropic nested Fermi surface.
Especially near half-filling, there are van Hove singularities in
addition to non-uniformity of the Fermi velocities along the
Fermi surface. Moreover, in the low-energy limit, the system is not a
regular Fermi liquid due to nesting. For instance, the
quasiparticles are expected to have a short lifetime, $1/\tau \sim T$,
near the Fermi surface,
and the existence of a well-defined Fermi surface may be
questioned.\cite{Virosztek}
To incorporate these characteristics, we introduce a low-energy
theory which models the band structure of the two-dimensional
Hubbard model near half-filling. We  make use of multidimensional
bosonization\cite{Haldane,Tony,Fradkin,Kopietz}
to describe the low-energy properties of the model. Although
Fermi liquid theory may break down near the Fermi surface,
marginal deviations from a Fermi liquid are within the scope of
bosonization.

The RG in the bosonic basis is a suitable method to study the
low-energy phases of the model
since it puts all interaction channels on an equal footing.
Wilson's RG algorithm, which we shall make use of, consists of integrating
out high-energy degrees of freedom to construct a low-energy effective
theory. In a degenerate fermion system, this amounts to successively
eliminating inner and outer sides of the momentum shell around the
Fermi surface in which the low-energy theory
lies.\cite{Shankar,Shankar1,Chitov}
In this paper we construct a low-energy effective theory of a nested
fermion liquid  by the RG procedure and then investigate the
RG flow of the effective interactions in the framework of bosonization.
In other words, we successively integrate out higher-energy modes of
the bosonic field $\phi$. In most cases, this bosonized RG procedure
yields results which agree with those obtained in the fermion basis.
The distinct advantage of the RG treatment based on bosonization is
that the existence of well-defined quasiparticles is not {\it a priori}
assumed. Furthermore, those interaction channels which
preserve the infinite-$U(1)$ symmetry\cite{Haldane} on the Fermi
surface are exactly diagonalized at the outset.\cite{HKM}

This paper is organized as follows.
In Sec. \ref{sec:model}, we introduce the model low-energy theory of
a two-dimensional nested Fermion liquid. We use multidimensional
bosonization
to incorporate the zero sound channel and show that the resulting
low-energy fixed point is a marginal Fermi liquid.
In Sec. \ref{sec:RGBCS}, we derive the RG equations for the BCS channel
and discuss
the condition of the BCS instability. An analog of the Landau
theorem is applicable to the case of anisotropic Fermi surfaces.
In Sec. \ref{sec:RGRI}, we incorporate marginally relevant repulsive or
AF interactions and evaluate their RG flow which contribute to the BCS
channel. The $d_{x^2-y^2}$-wave channel is the most important one,
and we argue that below half-filling competition arises between the
spin-density-wave (SDW) and $d_{x^2-y^2}$-wave superconductivity.
A brief summary and conclusions are given in Sec. \ref{sec:NestConcl}.

\section{Model Low-energy Theory}
\label{sec:model}

The Hubbard model is a convenient description of
strongly correlated fermion systems.
For a reliable investigation into the RG of this model,
it is essential to accommodate a properly realistic
band structure near the Fermi surface.
Here we present a model low-energy theory which
 simulates the nested fermion
system.

First, consider the band structure of the
Hubbard model in two dimensions near half-filling.
The Hamiltonian is
\begin{equation}
H=-t\sum_{i,j} c^{\dag \alpha}_i c_{j \alpha}
+U\sum_i n_{i \uparrow}
{}~n_{i \downarrow}~,
\label{hubbard}
\end{equation}
where $ i,j$ are between nearest neighbors on a square lattice
 and $U$ is the on-site
interaction strength.
Here $c^{\dag \alpha}_i$ and $ c_{i \alpha}$ are fermion creation and
annihilation operators.
Throughout this paper we assume that $|U/t| \ll 1$ holds so that the
perturbative expansions are justified.
For now we turn off the on-site interaction term and discuss
the character of the Fermi surface that results from the hopping term.
The spectrum of the fermions is
\begin{equation}
E({\bf k})=-2t (\cos k_x a +\cos k_y a ),
\label{spec}
\end{equation}
where $a$ is the lattice spacing.
At half-filling,
the Fermi surface is described by the points in $k$ space which
satisfy
\begin{equation}
 \cos k_x a + \cos k_y a =0 ~.
\label{trace}
\end{equation}
The Fermi velocity is a function of $\bf k$ given by
\begin{equation}
 {\bf v(k)} = 2ta ({\bf \hat{x}}~\sin (k_x a) +{\bf \hat{y}}~
\sin (k_y a) )
\label{vel}
\end{equation}
 on the Fermi surface.
Singularities in the density of states occur at four corners of
the Fermi surface, ${\bf k}=(\pm \pi /a,0)$
or $(0,\pm \pi /a)$, where the Fermi velocity vanishes.

Turning on the on-site interaction, the RG may be used
 to integrate out the high-energy
degrees of freedom.\cite{Shankar}
The resulting low-energy effective theory
is expressed in terms of quasiparticles living in a narrow shell
around the Fermi surface of energy cutoff $\varepsilon_c $.
Quasiparticles may then be related to the bare fermion operators
by
\begin{equation}
\psi_{\bf k}=Z_{\bf k}^{-1/2}(\varepsilon_c )~c_{\bf k} ~,
\end{equation}
where the wave-function renormalization factor
$Z_{\bf k}(\varepsilon_c)$
rescales the discontinuity in the quasiparticle occupancy
at the Fermi surface back to 1. The final quasiparticle weight
$Z_{\bf k}$ can be obtained by further integrating out the remaining
low-energy degrees of freedom,
$\lim_{\varepsilon_c \to 0} Z_{\bf k}(\varepsilon_c)
=Z_{\bf k}$. In non-Fermi liquids, $Z_{\bf k}$ is zero.
Here we stop the RG process at the energy scale $\varepsilon_c$
and investigate the low-energy theory thus obtained.
Assuming that the
weak nearest-neighbor interaction
term does not alter the shape of the Fermi surface
or the band structure in  Eq. (\ref{spec}), we may linearize the
spectrum of the quasiparticles:
\begin{equation}
E({\bf k})=v~|\sin [k_F({\bf k})_x a] |~q_{\parallel}({\bf k})~,
\label{renspec}
\end{equation}
which is similar to Eq. (\ref{spec}) close to the Fermi surface.
 Here $q_{\parallel}({\bf k})\equiv {\bf (k-k_F(k))\cdot
\hat{n}(k)}$, where $\bf k_F(k)$ is the wave-vector on the
Fermi surface closest to $\bf k$,  $k_F({\bf k})_x$ is the
$x$ component of $\bf k_F(k)$, and $\bf \hat{n}(k)$ is the
unit vector normal to the Fermi surface at $\bf k_F(k)$.

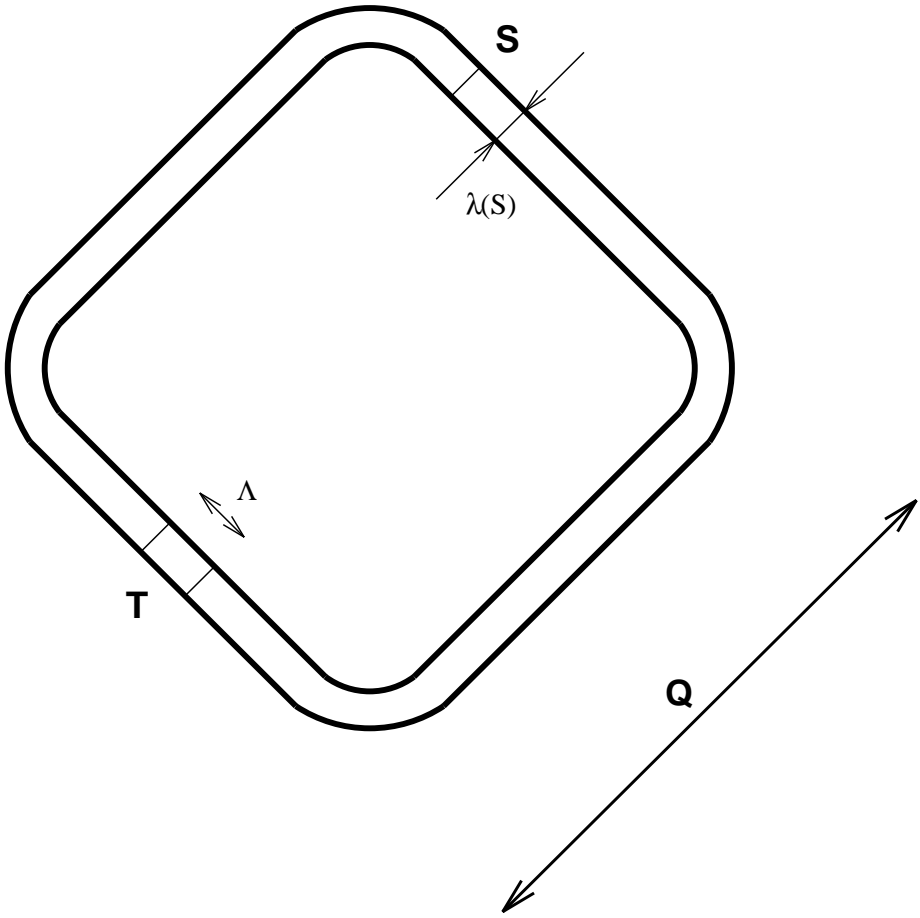
\begin{figure}
\caption{The low-energy effective theory in the vicinity of the Fermi
surface. The thin momentum shell is coarse-grained into squat boxes of
a height $\lambda$ and a width $\Lambda$. $\bf Q$ is the nesting
vector.}
\label{fig:nest1}
\end{figure}

The difficulty with bosonizing this Fermi surface is that at the
four corners of the Fermi surface the Fermi velocities are zero
and the Fermi surface normal vectors are not well-defined in the
vicinity of the van Hove points.
Below half-filling, however,
the van Hove singularity is not encountered.
So we introduce a model spectrum which simulates the Hubbard model
near half-filling but with a smoothed out van Hove singularity.
We modify the spectrum in Eq. (\ref{renspec}) as 
\begin{eqnarray}
E({\bf k})&=&v~|\sin [k_F({\bf k})_x a]|~q_{\parallel}({\bf k})~
,~\epsilon /a<|k_F({\bf k})_x|
<(\pi -\epsilon)/a ~, \nonumber \\
&=&v\sin (\epsilon a)~q_{\parallel}({\bf k})~~{\rm (otherwise)}
\label{modspec}
\end{eqnarray}
and we smooth the four corners of the Fermi surface as four
quarter-circles with a radius $\sqrt{2} \epsilon/a$ (Fig. \ref{fig:nest1}).
So we define the Fermi velocity function $v_F({\bf k})$
as the coefficient of $ q_{\parallel}(\bf k)$ in Eq. (\ref{modspec}).
This choice is reasonable because the number of states near
the van Hove points here is of order $\epsilon /\sin(\epsilon)
\sim O(1)$, which is finite even if the limit $\epsilon \rightarrow 0$
is taken. This artificial
smoothing simulates the van Hove singularity
as the van Hove points accommodate a fixed fraction of the
total number of states.
{}From now on we set $1/a =1$ and $v =1$ for convenience.
The factor $1/a$ may be recovered whenever a comparison with
other momentum scales
is needed. With this convention, for example, the nesting
vectors are expressed as
 ${\bf Q} \equiv \pm (\pi ,\pi )$ or $\pm (\pi ,-\pi)$.

Now that the Fermi velocities are well-defined over the
whole Fermi surface, the model may be bosonized.
We coarse-grain the Fermi surface into small squat boxes
of a width $\Lambda $ along the Fermi surface
and a height $\lambda(S)\equiv \varepsilon_c /v_F(S)$
normal to the Fermi surface, as shown in
Fig. \ref{fig:nest1}. Here $v_F(S)$ is the Fermi velocity
function at the squat box $S$.
Then we introduce a coarse-grained charge current which
lives in a squat box labeled as $S$:
\begin{equation}
J_{\alpha}({\bf k_S;q})=\sum_{\bf k}\theta({\bf S;k-q})
{}~\theta({\bf S;k})~\{ \psi^{\dag}_{{\bf k-q},\alpha}~
\psi_{{\bf k},\alpha}
-\delta_{\bf q,0}~n_{{\bf k},\alpha} \}~.
\label{defJ}
\end{equation}
Here $n_{{\bf k},\alpha}=\langle \psi^{\dag}_{{\bf k},\alpha}
\psi_{{\bf k},\alpha}
\rangle$. $\alpha $ is the spin index and the label $S$ identifies
a patch on the Fermi surface with a Fermi momentum $\bf k_S$
and $\theta({\bf S;k})=1$ if $\bf k$ lies inside a
squat box centered on $\bf k_S$ and $\theta({\bf S;k})=0$
otherwise. The two momentum scales $\Lambda, \lambda(S) $
are taken to be small
so that $1/a \gg \Lambda \gg \lambda(S) $ at each Fermi surface
patch. The latter
inequality is needed to minimize scattering of fermions outside
of the squat box so that the number of fermions in each patch
on the Fermi surface is conserved.
Now the limit of
$\epsilon \rightarrow 0$ is not permitted since it violates the condition
of $ \epsilon /a \gg \Lambda $ at the four van Hove points.
Instead, we take $\epsilon $ as a small non-zero number.

The $U(1)$ current algebra reads\cite{Haldane,Tony,Fradkin}
\begin{equation}
[J_{\alpha}({\bf k_S;q}),J_{\beta}({\bf k_T;p})]=
\delta_{\alpha \beta}
\delta_{S,T}\delta^2_{\bf q+p,0}\Omega~{\bf q\cdot \hat{
n_S}}~,
\label{U1}
\end{equation}
which holds when $1/a \gg \Lambda \gg \lambda(S) $.
Here $\Omega \equiv \Lambda (L/2\pi )^2$ is the number
of states in the squat box divided by $\lambda (S)$ and
$\bf \hat{n_S}$ is a unit vector normal to the Fermi surface at
patch $S$.
The algebra, Eq. (\ref{U1}), makes it possible to bosonize
the currents.
The bosonic representation of the currents
is
\begin{equation}
J_{\alpha}({\bf k_S;x})=\sqrt{4\pi}~{\bf \hat{n_S}\cdot
\nabla}\phi_{\alpha}({\bf k_S;x})~,
\end{equation}
where the bosonic fields $\phi_{\alpha}$
satisfy the canonical commutation relations.
Then the fermion quasiparticle fields at each patch
can be expressed in terms of boson fields as\cite{Tony}
\begin{eqnarray}
\psi_{\alpha}({\bf k_S;x},t) &\equiv&
e^{i\bf k_S\cdot x}\sum_{\bf p} \theta({\bf S;p}) e^{i\bf (p-k_S)
\cdot x} \psi_{\alpha \bf p}(t) \nonumber \\
&=&\sqrt{\Lambda \lambda (S)/2\over
(2\pi)^2}~e^{i{\bf k_S \cdot x}} \exp \Big{\{}
i {\sqrt{4\pi}\over \Omega}\phi_{\alpha}({\bf k_S;x},t)
\Big{\}}~\hat{O}(S)~.
\label{bosonize}
\end{eqnarray}
Here $\hat{O}(S)$ is an ordering operator introduced
to maintain Fermi statistics along the Fermi surface.

The bosonized effective action of the low-energy theory
is
\begin{eqnarray}
S &=& \sum_S \sum_{{\bf q},\lambda (S)>{\bf q\cdot \hat{n}_S}>0}
\int {d\omega \over 2\pi } [\omega -v_F(S){\bf q\cdot \hat{n}_S}]~
a^{*\alpha }({\bf k_S};q)~a_{\alpha }({\bf k_S};q)
+S_I ~.
\end{eqnarray}
Here $S_I$ is the fermion interaction terms which we
discuss later.
The charge currents are related to the canonical boson operators
$a$ and $a^{\dag}$ by
\begin{eqnarray}
J_{\alpha}({\bf k_S;q})=\sqrt{\Omega |{\bf q\cdot \hat{n}_S}|}
{}~[a_{\alpha}({\bf k_S;q})~\theta ({\bf q\cdot \hat{n}_S})
+a^{\dag}_{\alpha}({\bf k_S;-q})~\theta (-{\bf q\cdot \hat{n}_S})
]~,
\end{eqnarray}
where
\begin{equation}
[a_{\alpha}({\bf k_S;q}),~a^{\dag}_{\beta}({\bf k_T;p})]
=\delta _{\alpha \beta }\delta _{S,T} \delta ^2_{\bf p,q} ~.
\end{equation}
Here $\theta(x)=1$ if $x>0$ and $\theta (x)=0$ otherwise.
{}From Eq. (\ref{bosonize}) the fermion Green's function
can be expressed in terms of the Fourier-transformed boson correlation
function and it is given by\cite{HKM}
\begin{eqnarray}
G^{\alpha}_F({\bf k_S;x},t)&\equiv & i\langle \psi ^{\dag \alpha }
({\bf k_S;x},t)~\psi _{\alpha }({\bf k_S;0},0) \rangle
\nonumber \\
&=&{\lambda(S) \Lambda /2\over (2\pi )^2}e^{i{\bf k_S \cdot x}}
\exp \Big{[} \int {d^2q \over (2\pi)^2} \int {d\omega \over 2\pi}
[e^{i({\bf q \cdot x}-\omega t)}-1] \nonumber \\
&&\times {(2\pi)^2\over \Lambda {\bf q\cdot \hat{n}_S}}
\langle a_{\alpha}({\bf k_S;q}) ~a^{\dag}_{\alpha}({\bf k_S;q})\rangle
\Big{]}~.
\label{GF}
\end{eqnarray}

Next consider the renormalization of the fermion interactions.
As in the case of a circular Fermi surface, the marginal interactions
take the form of four-fermion (two-body) interactions, as can be
verified by counting the engineering dimension.
Unlike the case of a circular Fermi surface, however,
many large-momentum scattering channels are now marginal due to the
geometry of the nesting. Some of the channels  give rise to
spin-density-wave (SDW) or charge-density-wave (CDW) instabilities.
These logarithmically renormalized channels are discussed
in Sec. \ref{sec:RGRI}.
For now, the zero-sound (ZS) channel may be incorporated.
This channel yields a Fermi liquid solution for a circular Fermi surface
in two dimensions and a Luttinger liquid in one dimension.
It is important to include this
channel in the free action before
 the the large-momentum
scattering channels are studied perturbatively.
This permits a check on whether the
Fermi surface breaks down in the presence of the ZS channel.

For simplicity, we suppress all momentum dependence in the coupling function
and drop the spin indices. Inclusion of the spin exchange channel does
not change the result qualitatively.
The ZS channel in the fermion basis is 
\begin{equation}
S_{ZS}=-\int d^2x~dt~F(S,T)~\psi^{\dag}({\bf k_S;x},t)~
\psi({\bf k_S;x},t)~\psi^{\dag}({\bf k_T;x},t)~\psi({\bf k_T;x},t)~.
\end{equation}
Note that a $U(1)$ phase rotation in each squat box,
$\psi({\bf k_S;x},t) \rightarrow e^{i\Gamma(S)}\psi({\bf k_S;x},t)$
and $\psi^{\dag}({\bf k_S;x},t) \rightarrow e^{-i\Gamma(S)}
\psi^{\dag}({\bf k_S;x},t)$, leaves the action invariant.
Physically, this infinite $U(1)$ symmetry means that the number of
fermions at each patch on the Fermi surface is conserved.
This channel can be expressed in terms of $U(1)$ currents
since  Eq. (\ref{defJ}) is $U(1)$ invariant at each patch.
The ZS channel in the bosonized form is succinctly written as
\begin{equation}
S_{ZS} =-{1\over V}\sum_{S,T}\sum_{\bf q}\int {d\omega \over 2 \pi }
{F(S,T)}~J({\bf k_S};q)~J({\bf k_T};-q)~,
\end{equation}
which is marginal since it is bilinear in bosons.
Here $q$ denotes $({\bf q},\omega)$.
The solution is a Gaussian
boson parametrized by the shape of the Fermi surface and the coupling
parameter $F(S,T)$.
Those which do not have infinite $U(1)$ symmetry are not
expressed as bilinear in currents but they can be studied perturbatively.
Here for simplicity the case $F(S,T)=F_0$ is studied.

Turning our attention to the fermions on the 
nested part of the Fermi surface, since
that is where the most singular behavior may arise,
we make use of the generating functional method,\cite{KHM,KHM2}
and obtain the
boson correlation function
\begin{eqnarray}
\langle a({\bf k_S};q) a^{\dag}({\bf k_S};q) \rangle
&=& {i\over \omega - v_F(S){\bf q\cdot \hat{n}_S}}+
i{\Lambda \over (2\pi )^2}~{{\bf q\cdot \hat{n}_S}\over
(\omega - v_F(S){\bf q\cdot \hat{n}_S})^2}~\big{[} K_0(q) \big{]} ^{-1}
{}~,
\label{aac}
\end{eqnarray}
and the fermion Green's function is given by Eq. (\ref{GF}).
Here,
\begin{eqnarray}
K_0(q) &=& {1\over F_0}-{2\Lambda \over (2\pi )^2}\sum_S {\theta(
{\bf q\cdot \hat{n}_S}) ~{\bf q\cdot \hat{n}_S}\over
\omega - v_F(S){\bf q\cdot \hat{n}_S}} \nonumber \\
&=& {1\over F_0}+ {\sqrt{2} \over (2\pi )^2 }\Big{[}
(2\pi )^2~\chi_0(\omega, \sin \epsilon ~\bf q)
+\chi _N (\omega ,{\bf q}) \Big{]} ~,
\label{K0}
\end{eqnarray}
where $\chi_N$ is the contribution from the nested part of the Fermi
surface:
\begin{eqnarray}
\chi_N (q) &=& {4\over \sqrt{1-\omega ^2/
q_{\parallel} ^2}} \ln \bigg{[}{\sqrt{1-\omega ^2/q_{\parallel} ^2}
+\cos \epsilon \over\sqrt{1-\omega ^2/q_{\parallel} ^2}-\cos \epsilon }
\bigg{]} \nonumber \\
&&+{4\over \sqrt{1-\omega ^2/
q_{\perp} ^2}} \ln \bigg{[}{\sqrt{1-\omega ^2/q_{\perp} ^2}
+\cos \epsilon \over\sqrt{1-\omega ^2/q_{\perp} ^2}-\cos \epsilon }
\bigg{]} ~.
\label{chiN}
\end{eqnarray}
Again $q_{\parallel}$ is the component of $\bf q$ parallel with the
Fermi surface normal and $q_{\perp} $ is the perpendicular component.
The contribution from the curved part of the Fermi surface ($\chi_0$) in
Eq. (\ref{K0}) gives a self-energy correction
${\rm Im} ~\Sigma \sim \omega ^2 \ln \omega $ as in a Fermi liquid
system.\cite{HKM}
We perturbatively estimate the imaginary part of the quasiparticle
self-energy which originates from Eq. (\ref{chiN}).
The imaginary part of $\chi_N
(q)$ arises when $| \omega | < |q_{\parallel}| ~\sin \epsilon $
or $| \omega | < |q_{\perp}| ~\sin \epsilon $.
To second order in $F_0$, the boson correlation function is expanded as
\begin{eqnarray}
\langle a({\bf k_S};q) a^{\dag}({\bf k_S};q) \rangle
&\approx & {i\over \omega - v_F(S){\bf q\cdot \hat{n}_S}} \nonumber
\\ &&+
i{\Lambda \over (2\pi )^2}~{{\bf q\cdot \hat{n}_S}\over
(\omega - v_F(S){\bf q\cdot \hat{n}_S})^2}~\Big{(}
F_0 -F_0^2 {\sqrt{2} \over (2\pi )^2}~ \chi_N(q)
 \Big{)}~,
\label{sopt}
\end{eqnarray}
where the $\chi_0$ term is dropped. The first order term in $F_0$
renormalizes the Fermi velocity at the quasiparticle pole just as in
the Fermi liquid case, but the velocity correction is
$\delta v = F_0 \Lambda /k_F (2\pi)^2$, which is infinitesimal.
Spin-charge
velocity separation does not occur, as can be easily
shown when the spin indices and the exchange channel
are included in the effective action.
Inserting Eq. (\ref{sopt}) into Eq. (\ref{GF}), expanding the
exponential to  second order in $F_0$, and performing a
Fourier transform into $({\bf k},\omega )$ space yields the
imaginary correction to the quasiparticle self-energy.\cite{HKM}
Then $ {\rm Im} ~\Sigma ({\bf k},\omega ) \propto
\omega~F_0 ^2 \Lambda /k_F $, which is of a marginal Fermi liquid
form.\cite{MFL}
Similarly, corrections of higher orders in $F_0$ can be
estimated and we find that they also yield a marginal Fermi
liquid form.
This agrees with previous expectations by others who obtained
the inverse lifetime of quasiparticles, $1/\tau \sim T$,
as a function of the temperature.\cite{Virosztek}
Since the quasiparticle weight of a marginal Fermi liquid
vanishes at the Fermi surface as the inverse of a logarithm,
the Fermi liquid fixed point breaks down but the
breakdown of the Fermi surface is smooth enough for the bosonization
and the subsequent RG procedure to be applicable.\cite{KHM2}
Therefore we can confidently apply the RG method to analyze
large-momentum channels which do not preserve infinite
$U(1)$-symmetry.

One might wonder whether the marginal-Fermi-liquid-like behavior
of the nested fermion system
 has relevance to the linear-$T$ in-plane resistivity of
cuprate materials in the normal state.\cite{Takagi}
However, we only have considered
a simplified single-band Hubbard model with the ZS channel alone
whereas a three-band Hubbard model may be a better description of the
cuprate materials. For example, we have not incorporated in the
model the strong van Hove singularity which is present in the
cuprate materials.\cite{Dessau}
Furthermore, the $n$-doped ${\rm N}_{2-x}{\rm Ce}_x {\rm Cu 
O}_{4-\delta}$ (NCCO) material
exhibits $T^2$ resistivity even though photoemission studies
show a nested Fermi surface.\cite{King}
This suggests that nesting with the ZS alone
is not sufficient to account for the behavior of the
resistivity in the normal states
of cuprate superconductors.

In the following sections, we study large-momentum scattering channels
by the perturbative RG method.
We do not explicitly include the ZS channel at the
outset but we
assume that it contributes sub-leading corrections to RG equations.
For example, though bosons in different patches are  correlated via the
ZS channel, the correlation between fermions in different patches is
negligible:
\begin{equation}
\langle \psi^{\dag}({\bf k_S;x})~\psi({\bf k_T;0}) \rangle
\propto {1\over L} e^{ \langle \phi({\bf k_S;x}) \phi({\bf k_T;0})
\rangle }\rightarrow 0~,
\end{equation}
where $L$ is the system size.\cite{HKM}
Also the self-energy corrections are of order $\Lambda a \ll 1$
times smaller than the leading order.
So the leading RG equations we obtain
in the following sections are assumed to be
unchanged even if the ZS channel is considered.

\section{RG of the BCS channel}
\label{sec:RGBCS}

In this section we study the RG flow of the BCS channel for the nested Fermi
surface and find an analog of the Landau theorem.
As for the case of an isotropic circular Fermi surface,
the BCS channel in a nested square
Fermi surface flows logarithmically in scale length since the Fermi surface
is symmetric under reflection at the origin in momentum space.
On a circular Fermi surface, the Landau theorem\cite{Lifshitz}
 states that there occurs a
BCS instability if any one value of $V_l$ is negative where
$V_l$ is the $l$th orbital angular momentum component of the BCS vertex
function $V(\theta_1 ,\theta_2 )=\sum_l e^{il(\theta_1 -\theta_2)}
V_l$. Here $\theta_1 ,\theta_2 $ are the angular variables on the
Fermi surface in  momentum space. This theorem can be seen in the RG
equation for $V_l$ derived by Shankar,\cite{Shankar,HKM}
\begin{equation}
{dV_l \over d~\ln s}= -~{1\over 2\pi }~V_l^2~,
\end{equation}
the solution of which is
\begin{equation}
V_l(s)={V_l \over 1+{V_l\over 2\pi }\ln s }~,
\end{equation}
for each $l$. Thus, if any $V_l <0$, it eventually leads to instability.

In the case of the nested Fermi surface there is no rotational invariance in
the BCS vertex function, and so $V({\bf k_S,k_T}) $ must be
expanded in a double Fourier series.\cite{Baranov}
Consequently the Fourier coefficients are coupled in the resulting
RG equations, unlike in the case of a circular Fermi surface.
First we discuss the possible symmetry of the order parameter in a
square lattice.
The BCS vertex function has $D_4$ symmetry and there are five
sets of basis functions which generate the two-dimensional
function space to which the vertex function belongs.
The five sets are represented as
$A_1$ and $A_2$ ($s$-wave), $B_1$ and $B_2$ ($d$-wave),
and $E$ ($p$-wave).\cite{Baranov}
The basis functions in each set are following:
\begin{eqnarray}
A_1 &:&~\cos 2nk_x ~, \nonumber \\
A_2 &:&~{\rm sgn}(k_y)~\sin 2nk_x ~,\nonumber \\
B_1 &:&~\cos (2n+1)k_x ~,\nonumber \\
B_2 &:&~{\rm sgn}(k_y)~\sin (2n+1)k_x ~,\nonumber \\
E   &:&~a~\sin (n+1/2)k_x+b~{\rm sgn}(k_y)~\cos (n+1/2)k_x~,
\label{5groups}
\end{eqnarray}
where we have parametrized the basis functions with $k_x $ and
the sign of $k_y$ on the square Fermi surface which is  defined by
$\cos k_x +\cos k_y =0$.
In general the BCS vertex function  on a
square lattice can be expressed as a double Fourier expansion in terms
of these basis functions. Due to the $D_4$ symmetry of
the BCS vertex function, the expansion
does not contain crossed terms between basis functions of different symmetry
sets, and so the five sets are decoupled from one another in the RG
equation at the outset.

Now we inspect the RG equation of the BCS obtained in bosonization
language,\cite{HKM}
\begin{equation}
{d V({\bf k_S,k_T})\over d (\ln s)}= -{\Lambda \over (2\pi )^2}~\sum_U
{1\over v_F(U)}~V({\bf k_S,k_U})~V({\bf k_U,k_T})~.
\label{RGBCS2}
\end{equation}
Notice that there is a weight function of $1/v_F(U)$ in the sum.
To analyze the RG equation, we modify the basis functions in
Eq. (\ref{5groups}) so that they are orthonormal under a summation
over the Fermi surface
with the weight function $1/v_F(U)$. The modified basis functions are
given by
\begin{eqnarray}
a_1(n,{\bf k}) &\equiv & \bigg{[}{v_F({\bf k})\over 2\pi \sqrt{2}}
\bigg{]}^{1/2}~\cos 2nk_x~,~a_1(0,{\bf k})
\equiv \bigg{[}{v_F({\bf k})\over 4\pi \sqrt{2}}\bigg{]}^{1/2}~,
\nonumber \\
a_2(n,{\bf k}) &\equiv & {\rm sgn}(k_y)~\bigg{[}{v_F({\bf k})
\over 2\pi \sqrt{2}}\bigg{]}^{1/2}~\sin 2nk_x ~,\nonumber \\
b_1(n,{\bf k}) &\equiv & \bigg{[}{v_F({\bf k})
\over 2\pi \sqrt{2}}\bigg{]}^{1/2}~ \cos (2n+1)k_x ~,\nonumber \\
b_2(n,{\bf k}) &\equiv &{\rm sgn}(k_y)~\bigg{[}{v_F({\bf k})
\over 2\pi \sqrt{2}}\bigg{]}^{1/2} ~ \sin (2n+1)k_x ~,
\nonumber \\
e(n,{\bf k}) &\equiv &\bigg{[}{v_F({\bf k})
\over 2\pi \sqrt{2}}\bigg{]}^{1/2}
{}~\big{[} c_1~\sin (n+1/2)k_x+c_2~{\rm sgn}(k_y)~\cos (n+1/2)k_x
\big{]} ~.
\label{modbas}
\end{eqnarray}
The BCS vertex function can be expanded as
\begin{eqnarray}
V({\bf k_S,k_T}) &=&
\sum_{n,m} \Big{\{}
C_1^{nm}~a_1(n,{\bf k_S})~a_1(m,{\bf k_T})
+C_2^{nm}~a_2(n,{\bf k_S})~a_2(m,{\bf k_T}) \nonumber \\
&&+C_3^{nm}~b_1(n,{\bf k_S})~b_1(m,{\bf k_T})
+C_4^{nm}~b_2(n,{\bf k_S})~b_2(m,{\bf k_T}) \nonumber \\
&&+C_5^{nm}~e(n,{\bf k_S})~e(m,{\bf k_T})
\Big{\}}~.
\label{GFE}
\end{eqnarray}
This expansion is valid as long as ${\rm min}\Big{(} |v_F({\bf k})|
\Big{)} >0$ which
holds in the model considered here.
In fact, ${\rm min} \Big{(}|v_F({\bf k})|\Big{)} $ is zero
only at exactly half-filling.

Now insert Eq. (\ref{GFE}) into  Eq. (\ref{RGBCS2})
to obtain the RG equations of the Fourier coefficients:
\begin{equation}
{d C_i^{nm}\over d \ln s} = -~{1\over (2\pi )^2}
\sum_{l=0}^{\infty} C_i^{nl}~C_i^{lm}
\end{equation}
where $i=1,...,5$. Now the analog of the Landau theorem in the
nested fermion liquid can be obtained. Namely, the diagonal
Fourier coefficients obey the equations
\begin{equation}
{d C_i^{nn}\over d \ln s} = -~{1\over (2\pi )^2}
\sum_{l=0}^{\infty} |C_i^{nl}|^2~.
\end{equation}
Here  the diagonal coefficients decrease as the length scale increases.
Thus, if any one value of $C_i^{nn}$ starts out negative,
a BCS instability occurs. This is similar to the Landau theorem in an
isotropic circular Fermi surface.
This criterion of instability permits us to identify the possible symmetry
of the resulting order parameter, since the gap function is related to
the BCS vertex function by the following:
\begin{equation}
\Delta_{\bf k}= -\sum_{\bf k^{\prime}}V({\bf k,k^{\prime}})
{}~{\Delta_{\bf k^{\prime}} \over
2E({\bf k^{\prime}})}~,
\end{equation}
where $\bf k, k^{\prime}$ are within the thin shell around the Fermi surface
and $E({\bf k^{\prime}})$ is the spectrum of the low-lying excitation
in the superconducting state.
In the next section we apply this result to search for a possible
BCS instability in the presence of a repulsive interaction.

\section{RG of repulsive interactions}
\label{sec:RGRI}

In this section we discuss the RG flow of repulsive interactions.
We only consider those channels which flow logarithmically in length
scale $s$.
To simplify the analysis we take the coupling functions to be independent
of momentum.

There are three classes of interaction channels which are logarithmically
renormalized in the presence of nesting. The first class is the
 CDW or SDW channel. In case of a repulsive bare Hubbard
interaction, it is the SDW channel that leads to an instability.
This channel is shown to have little overlap with the BCS
channel and is practically decoupled from the BCS channel.
The second and third classes are  $Q=0$ and $Q=2k_F$ repulsive (or AF
exchange) channels which renormalize the BCS channel.
In the following two subsections we study the RG equations in the
SDW (CDW)  channel and in
the other repulsive (or AF) channel which renormalize the BCS and
calculate the Kohn-Luttinger effect.

\subsection{CDW and SDW channels}
\label{subsec:DW}

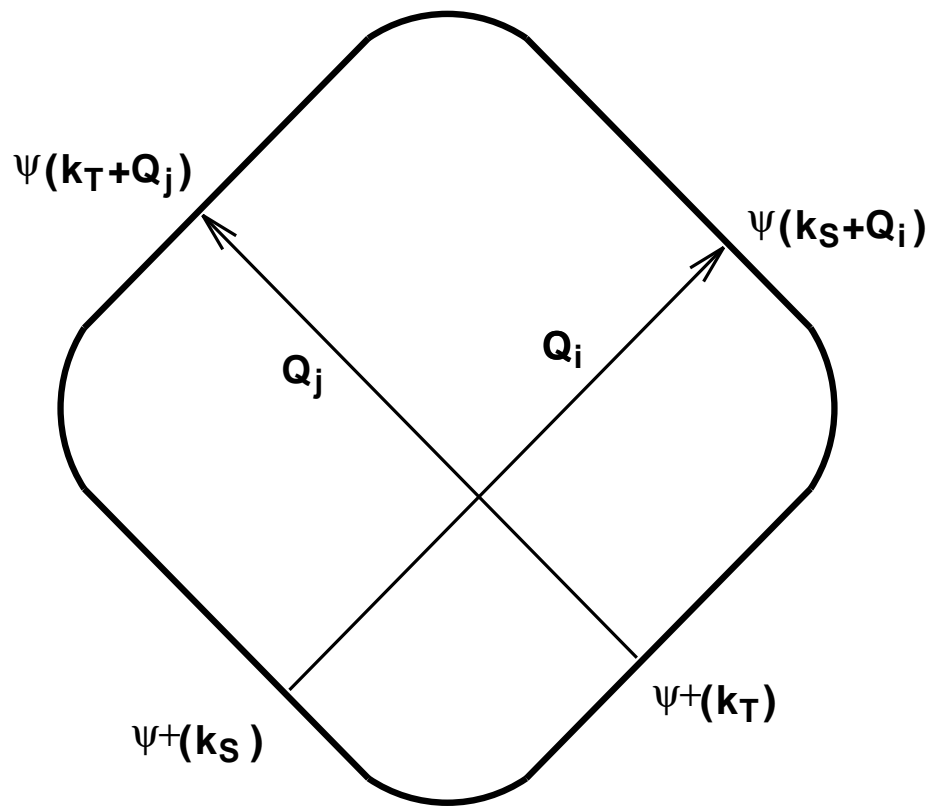
\begin{figure}
\caption{A typical density-wave scattering channel.}
\label{fig:nest2}
\end{figure}

In this subsection we derive the RG flows of density-wave channels,
which have negligible overlap with the BCS channel.
The channel is
\begin{eqnarray}
S_{DW} &=& \int d^2x~dt~{F_{DW}\over 2}\sum_{S,T} \sum_i
\psi^{\dag \alpha} ({\bf k_S};x)~ \psi_{\beta} ({\bf k_S+Q}_i;x)
\nonumber \\
&&\times \psi^{\dag \gamma} ({\bf k_T};x)~ \psi_{\delta} ({\bf k_T+Q}_j;x)~
[\delta ^{\beta}_{\alpha}\delta ^{\delta}_{\gamma}-\delta ^{\beta}
_{\gamma}\delta ^{\delta}_{\alpha} ]~.
\label{DW}
\end{eqnarray}
Here ${\bf Q}_i ,{\bf Q}_j$ are two of the four nesting vectors
(Fig. \ref{fig:nest2}).
Due to the nesting, this channel flows logarithmically in scale length
and causes a CDW or a SDW instability, depending on the sign of
$F_{DW}$.
The spin indices in Eq. (\ref{DW}) are appropriately
symmetrized according to  Fermi statistics.\cite{Chitov}
Separating the interaction $F_{DW}$ into decoupled
CDW and SDW channels,
\begin{eqnarray}
S_{DW} &= & {1\over 2}\int d^2 x~dt~
\sum^{\prime}_{S,T} \big{\{} F_{SDW}~{\bf S}_Q(S;{ x})\cdot
{\bf S}_Q(T;{ x})
+F_{CDW}~n_Q(S;{\bf x})~n_Q(T;{\bf x}) \big{\}}~,
\label{Hnest}
\end{eqnarray}
where
\begin{equation}
{\bf S}_Q(S;{\bf x})= {1\over 2}~\psi^{\dag \alpha}({\bf k_S;x})~
{\bf \sigma}^{\beta}_{\alpha} ~\psi_{\beta}({\bf k_S+Q;x})
\end{equation}
and
\begin{equation}
n_Q(S;{\bf x}) =\psi^{\dag \alpha}({\bf k_S;x})~
\psi_{\alpha}({\bf k_S+Q;x})
\end{equation}
are the spin density and the charge density, respectively.
The primed summation over $S,T$ indicates
that the sum is over the nested parts of the
Fermi surface.
First, we investigate the RG flow of the CDW term.
For simplicity we drop the spin indices.
{}From Eq. (\ref{bosonize}) the bosonized form of the CDW action is 
\begin{eqnarray}
S_{CDW} &=& {1\over 2} \sum_{S,T}^\prime
\int dt~d^2 x~ {F_{CDW} (S,T) }~
 \psi ^{\dag } ({\bf k_S};x)~\psi ({\bf k_S+Q}_i;x)~
\psi ^{\dag} ({\bf k_T};x)~
\psi ({\bf k_T+Q}_j ;x) \nonumber \\
&=& {1\over 2}\big{(}{ \Lambda \over (2\pi)^2}\big{)}^2
 \sum_{S,T}^\prime \int dt ~d^2 x~ {F_{CDW} (S,T)}~
{\lambda(S)~\lambda(T)\over 4}
\nonumber \\&&\times \exp \Big{\{}i {\sqrt{4\pi} \over \Omega}
[ \phi ({\bf k_S+Q}_i;x)+\phi ({\bf k_T+Q}_j;x)
-\phi ({\bf k_S};x)
-\phi ({\bf k_T};x) ]
\Big{\}}~ ,
\label{SU}
\end{eqnarray}
where  $x$ denotes
$({\bf x}, t)$.
The scaling  of $\varepsilon_c ~, ~x~,~t$ follows
\begin{equation}
t\rightarrow t^\prime /s ~,~x_{\parallel} \rightarrow
x^{\prime}_{\parallel} /s~,~\varepsilon_c \rightarrow
s\varepsilon_c ^\prime ~,
\label{scalelaw}
\end{equation}
where $s$ is the scale of the length. Dimensional analysis
shows that $S_{CDW} $ is scale invariant and thus marginal.

Next the fields may be split into higher and lower-energy modes
$\phi({\bf k_S};x) \rightarrow \phi ^\prime ({\bf k_S};x) +
 h({\bf k_S};x)$ where $h$ is the field
of higher-energy modes whose momentum ranges
$ \lambda ^\prime (S) /2 <
|{\bf p}| < \lambda(S) /2$ where $\lambda^\prime (S)\equiv
\varepsilon_c^\prime /v_F(S)$.
The equal-time correlation function of $h$ fields is
\begin{eqnarray}
\langle h({\bf k_S;x})~h({\bf k_T;0})\rangle
&=& \delta_{S,T}~{\Omega ^2\over 4\pi}~\ln \bigg{(}
{{\bf \hat{n}_S\cdot x}+2is/\lambda(S)\over
{\bf \hat{n}_S\cdot x}+2i/\lambda(S)}\bigg{)}~;~
|{\bf \hat{n}_S\times x}| \ll 1 \nonumber \\
&=& 0~;~ |{\bf \hat{n}_S\times x}| \gg 1~.
\label{hcorr}
\end{eqnarray}
We study how the partition function is renormalized after the fast
momentum modes are integrated out:
\begin{eqnarray}
\langle e^{- S_{CDW}} \rangle &=& \langle 1- S_{CDW} +{1\over 2}
S_{CDW}^2 +...\rangle \nonumber \\
&=& \exp \big{[} -S_{CDW}^\prime \big{]}~.
\label{expsu}
\end{eqnarray}
Here the
average in the brackets is taken only over the high-energy fields $h$
and $S_{CDW}^\prime $ is the renormalized CDW action.
The third term on the right hand side (RHS) of Eq. (\ref{expsu})
gives the second order in the $F_{CDW}$ correction to $S_{CDW}$.
Using the formula $\langle e^A~e^B\rangle =e^{\langle
AB+(A^2+B^2)/2\rangle }$ and Eq. (\ref{hcorr}), we obtain
\begin{eqnarray}
&&{1\over 2}\Big{[}\langle S_{CDW} ^2\rangle -\langle S_{CDW}\rangle ^2
\Big{]}
= {1\over 2}\bigg{[}{ \Lambda  \over (2\pi )^2}\bigg{]}^4
\sum^{\prime}_{S,T,U}\int dt~d^2x
{}~F^2_{CDW}~ {\lambda(S)~\lambda(T)~\lambda^2(U)\over 16s^4} \nonumber \\
&& \times
\exp \Big{\{} i{\sqrt{4\pi} \over \Omega}
[ \phi ({\bf k_S+Q}_i;x)+\phi ({\bf k_T+Q}_j;x)
-\phi ({\bf k_S};x)-\phi ({\bf k_T};x) ]
\Big{\}} \nonumber \\
&&\times\int du~d^2y~ \exp \Big{\{} {{4\pi}\over \Omega ^2}
\big{[}\langle h({\bf k_U};{\bf y},u)~h({\bf k_U};{\bf 0},u)\rangle
+\langle h({\bf k_U+Q}_l ;{\bf y},u)~h({\bf k_U+Q}_l ;{\bf 0},u)
\rangle \big{]}\Big{\}} \nonumber \\
&&+{\rm subleading ~terms} \nonumber \\
&\approx & {1\over 2}\bigg{[}{ \Lambda \over (2\pi)^2}\bigg{]}^4
\sum ^{\prime}_{S,T}\int dt~d^2x~
\sum^{\prime}_U F^2_{CDW}~{\lambda^2(U)\over
4s^2}
{}~{2\pi \over \Lambda}\int du~d({\bf \hat{n}_U\cdot y})~
{4s^2/\lambda(U) ^2
\over [v_F(U) u ]^2+({\bf \hat{n}_U\cdot y})^2} \nonumber \\
&&\times {\lambda(S)~\lambda(T)\over 4s^2}
{}~\exp\Big{\{} i{\sqrt{4\pi} \over \Omega}
[ \phi ({\bf k_S+Q}_i;x)+\phi ({\bf k_T+Q}_j;x)-\phi ({\bf k_S};x)
-\phi ({\bf k_T};x) ]
\Big{\}} \nonumber \\
&&+{\rm subleading~terms}~.
\end{eqnarray}
Here $v_F(U)$ denotes the Fermi velocity at the patch $U$.
Replacing $\lambda(S) $ with $s\lambda ^\prime(S)$
[$\lambda ^\prime(S)$ is
 the new cutoff] and performing the integrals
over $\bf \hat{n}_U\cdot y$ and $u$, we extract the
correction to the CDW coupling function
\begin{equation}
{d F_{CDW}(s)\over d \ln(s)}= - ~{\Lambda \over 2(2\pi )^2 }
\sum_U ^{\prime }{1\over v_F(U)}~F_{CDW}^2 ~.
\end{equation}
Therefore, we expect  instability in this channel if $F_{CDW}$
is strictly negative.
This agrees with the  result
 obtained by Shankar.\cite{Shankar}
$F_{CDW}<0$ corresponds to a bare attractive interaction $(U<0)$
and the CDW instability leads the phase transition into a
charge-density ordering with  period $(\pi ,\pi)$ and
$(\pi, -\pi)$.

Likewise, a similar RG transformation can be performed to obtain the
flow equation for the SDW channel:
\begin{eqnarray}
S_{SDW} &=& -~{1\over 2} \int d^2 x ~dt~\sum_{S,T} ^\prime F_{SDW}~
{\bf S}_Q  (S;x)\cdot {\bf S}_Q  (T;x)
\end{eqnarray}
yielding
\begin{equation}
{d F_{SDW}(s)\over d \ln(s)}= - ~{\Lambda \over (2\pi )^2 }
\sum_U ^{\prime }{1\over v_F(U)}~F_{SDW}^2
\label{RGFS}
\end{equation}
As expected there occurs an instability when $F_{SDW} <0$, which
corresponds to a bare repulsive interaction $(U>0)$.
The SDW instability leads to an AF spin-density ordering with
period $(\pi,\pi)$ and $(\pi,-\pi)$ since the on-site repulsive
interaction induces an AF exchange interaction with the nearest
neighbors.

Since the RG equation for the BCS channel is of interest, we
now discuss the overlap between the BCS and the SDW (CDW) channels.
In Fig. \ref{fig:nest3} we show a typical scattering channel which
 belongs to both the BCS and the density-wave channel.
In terms of the BCS, it can be
written as
\begin{eqnarray}
S_{\rm overlap} &=& \sum_{S,T} \int d^2 x~dt~V({\bf k_S,k_T})~
\psi ^{\dag \alpha}({\bf k_S};x)~\psi ^{\dag \beta}({\bf -k_S};x)~
\nonumber \\
&&\times \psi _{\beta}({\bf -k_T};x)~\psi _{\alpha}({\bf k_T};x)~
\delta _{\bf k_T,k_S+Q}~.
\end{eqnarray}
Obviously there is a severe phase space restriction due to the
factor of $\delta _{\bf k_T,k_S+Q}$ in the overlapped channel, which
amounts to a factor of $\Lambda a \ll 1$. Therefore, the SDW (CDW)
is practically decoupled from the BCS channel.
The renormalization of the BCS channel comes exclusively from the other
repulsive or AF exchange channels which we discuss in the next
subsection.
Here we can evaluate the energy scale of the SDW instability
for future reference.
{}From Eq. (\ref{RGFS}), we obtain the critical energy at which the SDW
instability occurs:
\begin{equation}
E_{SDW} \sim E_F ~\exp \bigg{[}- ~{1\over {8\sqrt{2}\over (2\pi)^2
}\ln {2\over \epsilon}~|F_{SDW}|} \bigg{]}~.
\label{ESDW}
\end{equation}

\subsection{Kohn-Luttinger effect}
\label{subsec:KL}

The two classes of repulsive channels which renormalize the BCS are
shown in Fig. \ref{fig:nest4} and Fig. \ref{fig:nest5}
and are given by
\begin{eqnarray}
S_R &=& {1\over 2}\sum_{S,T}\sum_{\bf \Delta p}\int d^2 x~dt~
F_R~\psi ^{\dag \alpha}({\bf k_S};x)~ \psi _{\beta}
({\bf k_S+\Delta p +Q}_i ;x) \nonumber \\
&&\times
\psi ^{\dag \gamma}({\bf k_T};x)~ \psi _{\delta}
({\bf k_T-\Delta p -Q}_i ;x)~[\delta ^{\beta}_{\alpha}
\delta ^{\delta}_{\gamma} -\delta ^{\beta}_{\gamma}
\delta ^{\delta}_{\alpha}] ~ ,
\end{eqnarray}
which is the zero momentum process (Fig. \ref{fig:nest4}) and
\begin{eqnarray}
S_U &=& {1\over 2}\sum_{S,T}\sum_{\bf \Delta p}\int d^2 x~dt~
F_U~\psi ^{\dag \alpha}({\bf k_S};x)~ \psi _{\beta}
({\bf k_S+\Delta p +Q}_i ;x) \nonumber \\
&&\times
\psi ^{\dag \gamma}({\bf k_T};x)~ \psi _{\delta}
({\bf k_T-\Delta p +Q}_i ;x)~[\delta ^{\beta}_{\alpha}
\delta ^{\delta}_{\gamma} -\delta ^{\beta}_{\gamma}
\delta ^{\delta}_{\alpha}] ~ ,
\end{eqnarray}
which is the $2k_F=Q$ process allowed because $2Q$ is a commensurate
wave-vector (Fig. \ref{fig:nest5}).
In both channels we take the coupling functions to be
momentum-independent and the spin indices are anti-symmetrized.\cite{Chitov}

Now we expand the partition function to second order in $S_R+S_U$,
integrate out the high-energy modes, and study the resulting
effective action generated by this procedure.
In the course of this RG procedure, $S_R$ and $S_U$ are renormalized
as well as the BCS channel, and so the resulting flow equation will be a
complicated set of coupled differential equations. We do not attempt to
perform a complete analysis of the flow equations
but we are only interested in
the trend of initial growth of the BCS channel; the higher-order
coupled equations
are important only in the very-long-length scale.
Thus we only study how $S_R+S_U$ renormalizes the BCS channel
in the initial stages of  growth near $s=1$.
The second-order correction to the BCS channel 
is calculated in the same manner as in the previous subsection:
\begin{eqnarray}
{1\over 2}\langle (S_R+S_U)^2 \rangle
&\approx& \ln s~( F_R^2 +4~F_U^2)\sum_S \sum_{\bf \Delta p}
\int d^2 x~dt~\psi^{\dag \alpha}({\bf k_S};x)~\psi^{\dag \beta}
({\bf -k_S};x)~\nonumber \\
&&\times
\psi_{\beta}({\bf k_S+\Delta p+Q};x)~
\psi_{\alpha}({\bf -k_S-\Delta p-Q};x)~\nonumber \\
&&\times {\sqrt{2}\over 2(2\pi )^2}~{1\over \cos \Delta p/2}
\ln \Big{(} {1+\cos (\epsilon +|\Delta p|/2)\over
1-\cos (\epsilon +|\Delta p|/2)} \Big{)}
\nonumber \\
&&+{\ln s\over 2}~F_R^2\sum_{S,T} \Big{[} {\sqrt{2}\over (2\pi )^2}
\ln \Big{(} {1+\cos \epsilon \over
1-\cos \epsilon } \Big{)} +{1\over 2\pi }\Big{]}
\nonumber \\
&&\times\int d^2 x~dt~\psi^{\dag \alpha}({\bf k_S};x)~\psi^{\dag \beta}
({\bf -k_S};x)~\psi_{\beta}({\bf -k_T};x)~
\psi_{\alpha}({\bf k_T};x)\nonumber \\
&&+~{\rm non-BCS~terms}~.
\label{crtn2}
\end{eqnarray}
The second term in Eq. (\ref{crtn2}) contributes to the
$s$-wave BCS channel. Since it is repulsive even in the
first order to begin with, the $s$-wave BCS channel is not a candidate for
instability.
The first term in Eq. (\ref{crtn2}) gives RG flow to all BCS
symmetry channels since it has a strong wave-vector dependence.
More succinctly, the RG equation of the BCS channel is given by
\begin{eqnarray}
{d V_{BCS}({\bf k_S,k_T})\over d \ln s} &=&
(F_R^2+4~F_U^2){\sqrt{2}\over 2(2\pi )^2}~{1\over \cos \Delta p(S,T)/2}
\ln \Big{(} {1+\cos (\epsilon +|\Delta p(S,T)|/2)\over
1-\cos (\epsilon +|\Delta p(S,T)|/2)} \Big{)}
\nonumber \\
&& -~{\Lambda \over (2\pi)^2}\sum_U {1\over v_F(U)}~
V_{BCS}({\bf k_S,k_U})~V_{BCS}({\bf k_U,k_T})~,
\label{flowEQ}
\end{eqnarray}
if $S$ and $T$ are in the same quadrant on the Fermi surface.
When $S$ and $T$ are in different quadrants,
\begin{equation}
{d V_{BCS}({\bf k_S,k_T})\over d \ln s}=-~{\Lambda \over (2\pi)^2}
\sum_U {1\over v_F(U)}~
V_{BCS}({\bf k_S,k_U})~V_{BCS}({\bf k_U,k_T})~.
\end{equation}
$\Delta p(S,T)$ is a vector along the nested Fermi surface which is shown
in Fig. \ref{fig:nest4}. For instance,
$\Delta p(S,T) = |{\bf (k_S-k_T)\cdot \hat{x}}-\pi|$ if
$\bf k_S$ and $\bf k_T$ are both in the first quadrant on the
$(k_x,k_y)$ plane. The expression of $\Delta p(S,T)$
in the other quadrants can be obtained by inverting the signs
of $\bf k_S$ or $\bf k_T$ appropriately.

Now we are ready to check which symmetry channels are attractive
by evaluating the generalized Fourier coefficients of the
double expansion shown in Sec. \ref{sec:RGBCS}.
Among the basis functions we introduced in Eq. (\ref{modbas}),
we especially focus on the fundamental modes whose Fourier
coefficients are presumably the most significant.
After performing the generalized Fourier transform of the RHS
of Eq. (\ref{flowEQ}) we find that there are two diagonal
Fourier coefficients which flow to negative numbers
among the fundamental modes.
Their rate of change  is evaluated as follows:
\begin{eqnarray}
{dC_2^{11}\over d \ln s} &=& \Lambda ^2\sum_{S,T}
{}~{1\over v_F({\bf k_S})~v_F({\bf k_T})}~
{dV_{BCS}({\bf k_S,k_T})\over d \ln s}~a_2(1,{\bf k_S})~a_2(1,{\bf k_T})
\nonumber \\
&\approx &-0.2~(F_R^2+4~F_U^2)-|C_2^{11}|^2/(2\pi)^2~,
\\
{dC_3^{00}\over d \ln s}&=& \Lambda ^2 \sum_{S,T}
{}~{1\over v_F({\bf k_S})~v_F({\bf k_T})}~
{dV_{BCS}({\bf k_S,k_T})\over d \ln s}~b_1(0,{\bf k_S})~b_1(0,{\bf k_T})
\nonumber \\
&\approx &-0.7~(F_R^2+4~F_U^2)-|C_3^{00}|^2/(2\pi)^2 ~,
\end{eqnarray}
where $C_2^{11}(s=1)= C_3^{00}(s=1)=0$ initially.
These are evaluated when $\epsilon \approx 0.1$.
Here we find that the inequality $\big{|}{dC_3^{00}\over d \ln s}\big{|}
>\big{|}{dC_2^{11}\over d \ln s}\big{|}$ holds for a sufficiently
small $\epsilon$.
{}From the Landau theorem we can infer that there can be a BCS instability
in either the $B_1$ ($d_{x^2-y^2}$-wave) or $A_2$ (eight-lobed $s$-wave)
channel as long as the SDW instability does not set in.
Furthermore, we can infer from the inequality
 $\big{|}{dC_3^{00}\over d \ln s}\big{|}
>\big{|}{dC_2^{11}\over d \ln s}\big{|}$ that
for a small enough $\epsilon$,  $d_{x^2-y^2}$-wave superconductivity
is preferred.
The reason the inequality holds is that the $d_{x^2-y^2}$-wave
basis function $b_1(0,{\bf k})$ has a larger overlap than
$a_2(1,{\bf k})$ does with the density
of states at the Fermi surface which peaks at the van Hove points.

Of course, $d_{x^2-y^2}$-wave superconductivity is not the only
possible ground state near half-filling. There is competition between
the SDW and  BCS instabilities since both are favored in the presence of a
bare on-site repulsive interaction.
Slightly below half-filling, the nesting  is not perfect but
nevertheless effective as long as the effective theory around the
Fermi surface covers a significant part of the nested Fermi surface.
As the energy cutoff around the Fermi surface is reduced via the RG
towards $\varepsilon_\delta$, where $\varepsilon_\delta$
is the energy deviation of the actual
Fermi surface from the nested one, the SDW and AF exchange channels stop
flowing whereas the BCS channel continues to be renormalized logarithmically
as its flow is independent of nesting.\cite{Shankar}
If initially the SDW coupling is small enough
or if $\varepsilon_\delta$ is large enough
so that it does not develop an instability
until $\varepsilon_c \rightarrow \varepsilon_\delta$,
or $E_{SDW}<\varepsilon_\delta$, there never occurs a SDW instability.
Although it is difficult to compute the critical energy,
the BCS instability eventually sets in at a sufficiently low-energy
scale.
{}From the argument above, the superconducting order parameter presumably has
 $d_{x^2-y^2}$-wave symmetry.
This result is consistent with the expectation that AF exchange
channel is conducive to  $d_{x^2-y^2}$-wave superconductivity
and qualitatively in good agreement with the parquet solution
obtained for a flat Fermi surface.\cite{parquet}

\section{Summary and conclusions}
\label{sec:NestConcl}

In this paper we investigated the BCS instability in
a nested Fermion liquid with a bare
repulsive interaction.
When the ZS channel was
incorporated we found that the low-energy fixed point was
of a marginal Fermi liquid type.
Although the Fermi liquid state broke down,
multidimensional bosonization was still
applicable.
Large-momentum scattering channels were then included
via the perturbative
RG procedure in the boson basis.
Among the logarithmically renormalized channels, the SDW channel showed the
quickest flow due to the nesting, but it had negligible overlap with
the BCS
channel. The other repulsive or AF channels gave rise to a
renormalization of the BCS couplings.
We found that the $d_{x^2-y^2}$-wave channel
was rendered the most attractive.
Close to but less than half-filling, when no instabilities
develop in the SDW channel ($E_{SDW}<\varepsilon_\delta$),
 a $d_{x^2-y^2}$-wave BCS instability sets in at sufficiently
low temperatures.
This is in agreement with the expectations of Monthoux, Balatsky,
and Pines,\cite{Monthoux} and the result obtained by Zhelenyak, Yakovenko,
and Dzyaloshinskii.\cite{parquet}

The procedure presented in this paper is useful for extracting
information about the low-energy fixed points of lattice systems.
Once a realistic band structure of the system near the Fermi surface
is given as the input, one can analyze the phase structure and the
order parameter symmetry at low temperatures.
As regards the cuprate superconductors, it would be
interesting to study the RG of a three-band Hubbard model. For a detailed
examination of the low-temperature phase structure of the materials,
numerical works on the RG equation are needed. In this manner,
one may be able to construct a phase diagram as a function of
temperature and the filling factor (or doping).

\section*{Acknowledgments}
The author thanks Youngjai Kiem, Brad Marston, and Jaejun Yu for
helpful comments and discussions and, in particular, Brad Marston
for a careful reading of the manuscript.

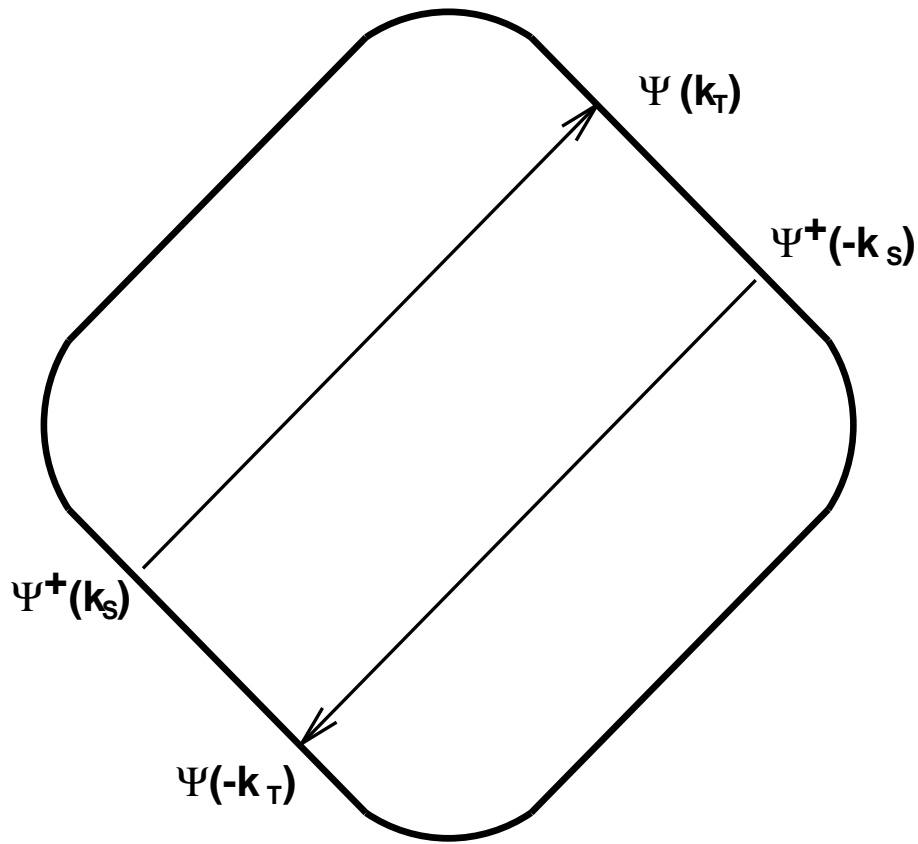
\begin{figure}
\caption{A typical overlap between density-wave and BCS channel.}
\label{fig:nest3}
\end{figure}
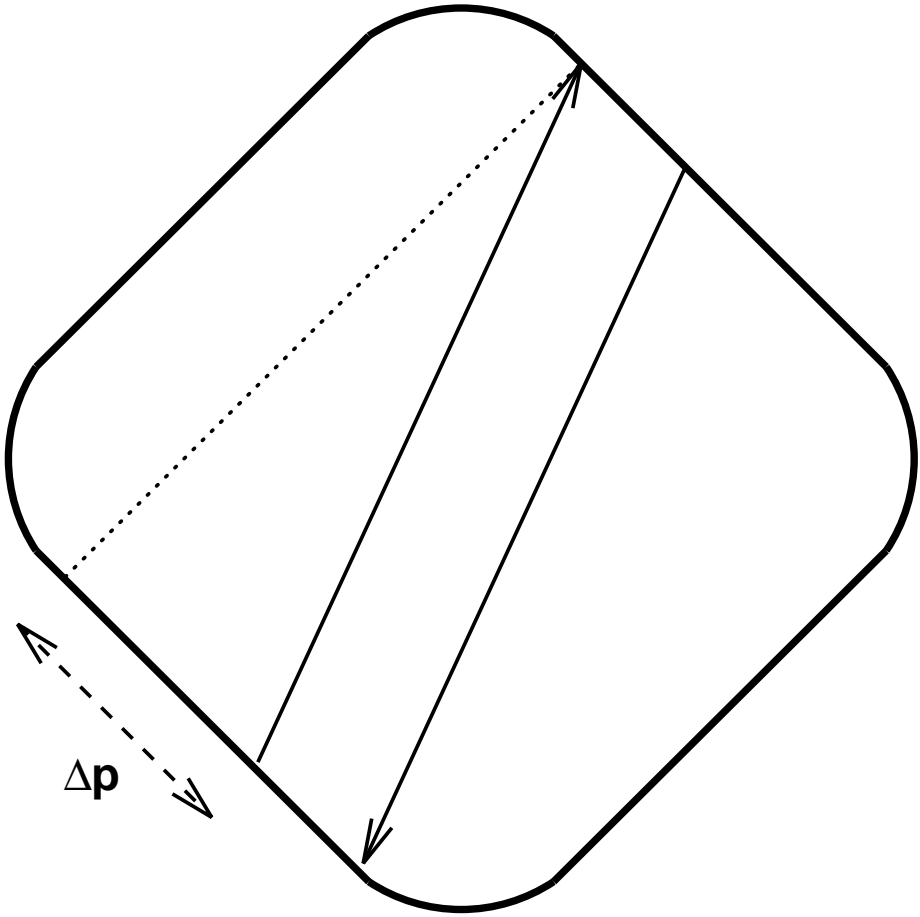
\begin{figure}
\caption{The repulsive or AF channels with a zero momentum
transfer which logarithmically renormalize
BCS channel.}
\label{fig:nest4}
\end{figure}
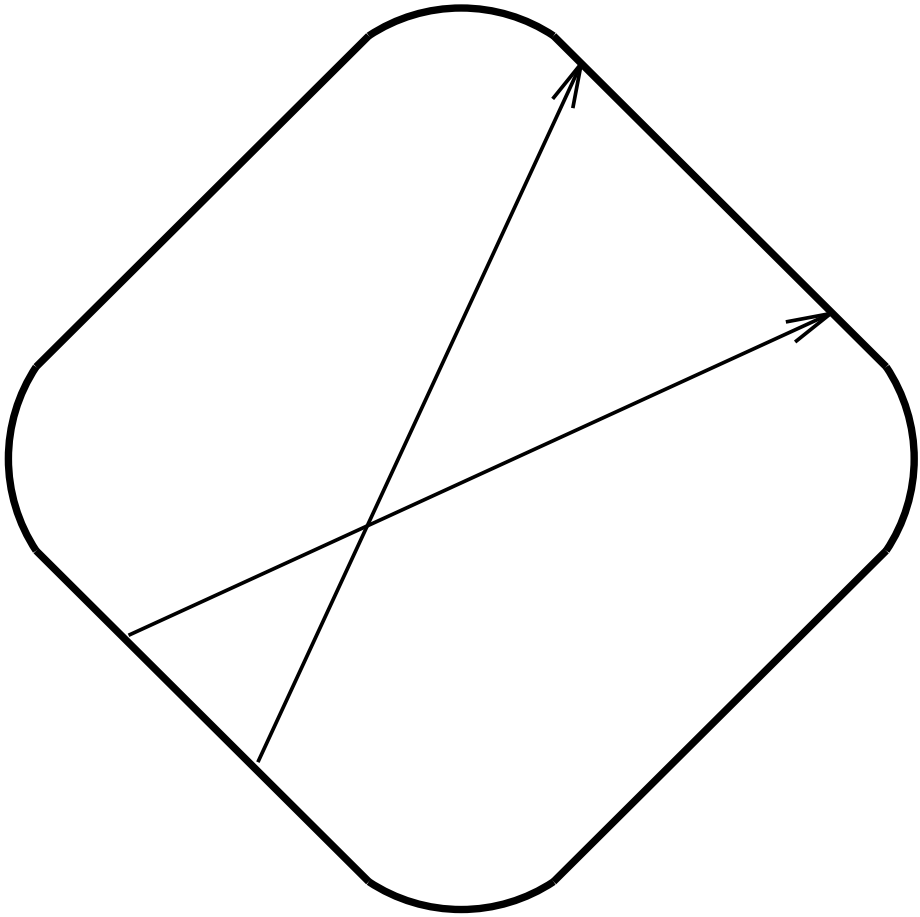
\begin{figure}
\caption{The repulsive or AF channels with a momentum
transfer $2Q$ which logarithmically renormalize
BCS channel.}
\label{fig:nest5}
\end{figure}

\end{document}